# Continuous-wave, single-frequency 229 nm laser source for laser cooling of cadmium atoms

YUSHI KANEDA,[1,3] J. M. YARBOROUGH,[1,3] YEVGENY MERZLYAK,[1,3] ATSUSHI YAMAGUCHI,[2,3] KEITARO HAYASHIDA,[2,4] NORIAKI OHMAE,[2,3,4] AND HIDETOSHI KATORI[2,3,4]

[1]College of Optical Sciences, University of Arizona, 1630 E University Blvd., Tucson, AZ, 85721
[2]Quantum Metrology Laboratory, RIKEN, Wako-shi, Saitama 351-0198, Japan
[3]Innovative Space-Time Project, ERATO, JST, Bunkyo-ku, Tokyo 113-8656, Japan
[4]Department of Applied Physics, Graduate School of Engineering, The University of Tokyo, Bunkyo-ku, Tokyo 113-8656, Japan
*Corresponding author: ykaneda@optics.arizona.edu

**Continuous-wave output at 229 nm for the application of laser cooling of Cd atoms was generated by the 4th harmonic using two successive second harmonic generation stages. Employing a single-frequency optically pumped semiconductor laser as a fundamental source, 0.56 W of output at 229 nm was observed with a 10-mm long, Brewster-cut BBO crystal in an external cavity with 1.62 W of 458 nm input. Conversion efficiency from 458 nm to 229 nm was more than 34%. By applying a tapered amplifier as a fundamental source, we demonstrated magneto-optical trapping of all stable Cd isotopes including isotopes $^{111}$Cd and $^{113}$Cd, which are applicable to optical lattice clocks.**

http://dx.doi.org/10.1364/OL.41.000705

High-power, continuous-wave (CW) deep ultraviolet (DUV) lasers have numerous industrial as well as scientific applications. They include material processing such as interferometric lithography, fiber Bragg grating fabrication, Raman sensing, defect inspection on semiconductor wafers and/or reticles, and spectroscopic applications [1, 2, 3]. For the conversion into this spectral region, the materials that can be used is fairly limited, as they must be transparent at the fundamental and harmonic, and able to phasematch for the particular combination of the wavelength. These available materials have nonlinear optical coefficients of just a few pm/V. With the exception of $CsLiB_6O_{10}$ (CLBO), which can phasematch for the harmonic wavelength of 238 nm and longer for second-harmonic generation (SHG), the effective nonlinear coefficient becomes smaller toward the shorter wavelength because of their crystal symmetry, making the generation of shorter wavelength increasingly challenging especially in CW.

In this Letter, we obtained 0.56 W of CW DUV power at 229 nm via direct SHG, without relying on more complex sum-frequency mixing normally used in this wavelength range. The feasibility and reliability of such CW DUV laser source was explored with its continuous operation of more than 100 hours. With the DUV laser source, we demonstrated magneto-optical trapping (MOT) of Cd atoms.

Because of their flexibility in their operating wavelength, optically pumped semiconductor lasers (OPSLs), also known as vertical external cavity surface emitting lasers (VECSELs) are suitable for application wherein specific wavelength is needed [4]. They combine the advantages of power scalability with good beam quality, and the flexibility in the operating wavelength, allowing efficient wavelength conversion. The externally resonant harmonic conversion of OPSLs is a suitable approach for the generation at a specific wavelength in the DUV region [5].

The laser device used in this experiment is InGaAs based device with upside-down configuration; the resonant periodic gain is grown on the GaAs wafer first, followed by the AlAs/GaAs high reflector. The laser resonator is formed by the high reflector on the OPSL device and an external concave output coupler with transmission of 1.5%, and 200 mm radius of curvature. The fundamental mode radius is 164 μm at the OPSL device, and the output end of the delivery fiber of the pump diode (810 nm) is imaged to ~166 μm radius on the device to match the $TEM_{00}$ mode of the OPSL resonator.

The OPSL device is mounted on a temperature-controlled base, which is maintained at 15°C throughout the experiment. Without any intra-cavity element, the laser can emit up to nearly 4 W in $TEM_{00}$ mode with 16 W of pump power. Increasing the pump power did not improve the output power due to thermal roll over in this configuration. An uncoated quartz plate birefringent filter at the Brewster's angle is inserted to coarse tune the wavelength, and a 0.8 mm-thick uncoated fused silica etalon is inserted to enforce single-frequency operation.  After the isolator, half-waveplate, phase modulator for Pound-Drever-Hall locking technique, and mode-matching lenses (spherical telescope), typically 3 W is available at 916 nm.

The first doubling cavity is a 4-mirror bow-tie ring cavity, and contains a 15-mm long lithium triborate ($LiB_3O_5$, LBO) crystal. LBO is cut for type-I phasematching at room temperature, ($\theta$ = 90°, $\varphi$ = 20.6°), and is antireflection coated on both sides at the fundamental and the second harmonic. The waist size of the cavity eigenmode at the fundamental at the center of the LBO crystal is ~32 × 28 µm, yielding the normalized conversion efficiency of $0.8 \times 10^{-4}$ W$^{-1}$.  The generated harmonic is out-coupled through one of the resonator mirrors.  With the fundamental power of 2.7 W, up to 1.62 W of harmonic at 458 nm is obtained. The efficiency is 60% for this stage, limited by the imperfect mode-matching of the fundamental into the first doubling cavity. It could have been better with more careful adjustment of mode-matching telescope, as well as the use of cylindrical lenses to mode-match the symmetric (round) beam of fundamental to the resonator mode having an aspect ratio of 1.23 (202 × 248 µm). Because of the walk-off in LBO crystal, the output beam at 458 nm is asymmetric. In order for mode-matching for the second doubling cavity, we used 3 cylindrical lenses.

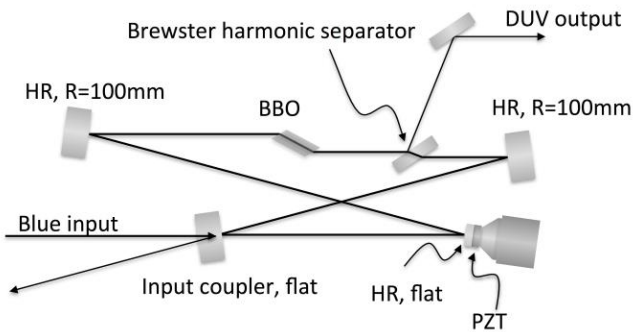

Fig. 1. Schematic of the 4th harmonic generation cavity.

Second doubling cavity is schematically shown in Fig. 1. Bow-tie ring cavity, with 7.5° of angle of incidence on 4 mirrors, contains a commercially available 10 mm-long beta-barium borate ($\beta$-BaB$_2$O$_4$; BBO) and a harmonic separator at the Brewster's angle, with its first surface coated for high reflectivity for the *s*-polarized harmonic and high transmission for *p*-polarized fundamental. Its second surface is uncoated, allowing low-loss transmission of the fundamental. Among the nonlinear materials that are transparent at 229 nm, BBO is currently the only practical choice that can phasematch for the harmonic generation at 229 nm and currently available in reasonable size with small optical loss for the fundamental at 458 nm.  At the phasematching angle of $\theta$ = 61.5° for type-I phasematching, BBO crystal has the effective nonlinear coefficient of 1.16 pm/V.  The cavity mode has a waist in the BBO crystal that is 46 × 37 µm, calculated from the resonator geometry, giving the normalized conversion efficiency of $0.62 \times 10^{-4}$ W$^{-1}$.  The flat input coupler is about 1.5% transmission at 458 nm. The resonator is contained in an airtight box and receives circulating, filtered clean dry air at a flow rate of 0.8 to 1.0 l/min in order to reduce the photo-contamination on the optical surfaces [6].

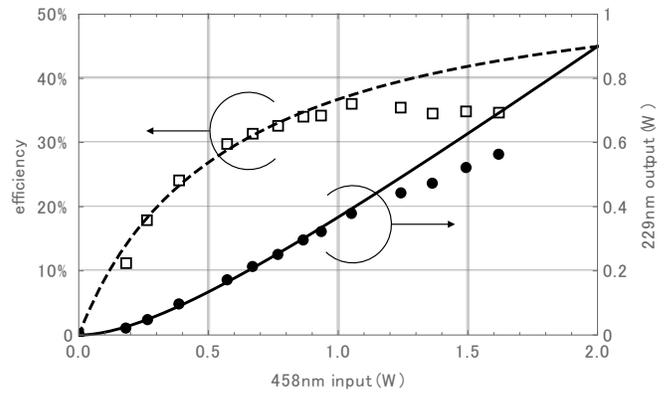

Fig. 2.  Input-output characteristic of the second doubling stage and the simulated result.

Figure 2 shows the input-output characteristic of the 4th harmonic generation as well as the simulation results. After the coarse adjustment of phasematching by the orientation, the temperature of the BBO was adjusted for optimum phasematching which was near 40°C, and was kept constant throughout the experiment.  With the input power of 1.62 W onto the cavity, a maximum output power of 0.56 W was observed. Conversion efficiency exceeded 34%. In the simulation, we assumed the mode-matching coefficient to be 80% and the resonator loss to be 0.3%.  Considering the reflection at the exit face of the BBO crystal for the *s*-polarized harmonic of 23%, the generated power at 229 nm was approximately 0.72 W. The internal efficiency, the generated 229 nm power by the total input power at 458 nm, exceeds 40%. As seen in Fig. 2, the conversion efficiency tended to saturate/reduce toward the higher input powers. We attempted to explain this by the under-coupling but it was not successful.  Therefore our explanation for this observation is the decrease in the normalized conversion efficiency, caused by the self-heating of the BBO crystal, giving non-uniform phasematching condition across the beam.  Considering the calculated temperature acceptance bandwidths of less than 2°C cm [7], small heating can very well affect the phasematch.

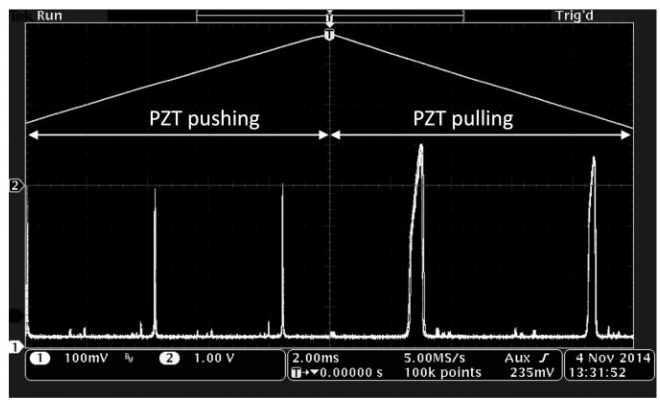

Fig. 3. Transmission from the 4th harmonic generation cavity with PZT scanned.

The indication of self-heating is also observed on the cavity transmission when the PZT on the cavity is scanned. When the PZT is linearly ramped, the self-heating of the BBO affects the sweep rate of the cavity length. With its negative thermo-optic coefficient ($dn/dT$), the self-heating slows down the sweep when the PZT is pulling, whereas it behaves the opposite when the PZT is pushing. The difference in width of the transmission peaks when the PZT is pulling and pushing indicates the self-heating as shown in Fig. 3. The circulating power at the resonance is tens of watts at 458 nm, therefore even a small amount of absorption, such as 0.1%/cm for example, in the mode volume of 46 × 37 μm, can cause the temperature rise of the order of degrees. In many of the BBO crystals we tested, such effect of self-heating is usually reduced when the phasematching is turned off by changing the temperature of BBO so that no significant DUV is generated. The majority of the self-heating is therefore caused by the DUV. However, in most of the BBO crystals, observable difference remained when DUV is not generated, so there was self-heating caused by the blue, too. This behavior is different from one crystal to another, sometimes among the pieces from the same manufacturer. We have not identified if the heating was from the surface or from the bulk material. With more careful and quantitative measurement of transmission peak width as well as its shape, it may be possible to evaluate the absorption of the crystals.

We operated the system over an extended period of time. We monitored the DUV output power, blue input power, and the photodetector output behind one of the high reflectors to represent the relative magnitude of internal power of the cavity. By dividing the DUV output power by the square of the internal power, we indirectly monitor the normalized conversion efficiency of BBO in relative scale. Figure 4 shows the power in DUV, efficiency from blue, and normalized conversion efficiency (in arb. unit) over a period of >120 hours. The conversion efficiency clearly drops over the course of operation, and the DUV power, which started from approximately 0.4 W drops to less than 0.2 W after about 120 hours, upon which we stopped the testing. Similar drop in the efficiency and output power would occur in typically less than 20 hours when there is no filtered circulating air to the unit, suggesting photo-contamination to be one of the causes of deterioration.

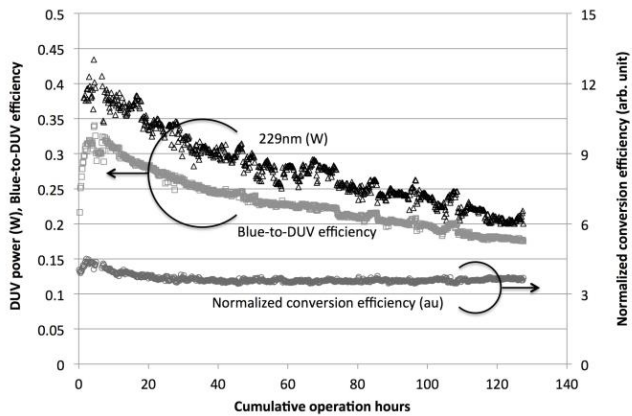

Fig. 4. DUV output power, efficiency from blue, and normalized efficiency over 120 hours.

Interestingly, the efficiency increases in the first few hours with the normalized conversion efficiency increasing in the same period. The normalized efficiency increases by about 10% in the first 3 - 4 hours, and starts to decrease until it settles to a steady value, which is about 10% less than the initial value, after approximately 30 hours, remaining at this value for the rest of the period. This behavior to some extent was observed in most of the samples we tested. We do not have good explanation on this. Varying the temperature of the crystal did not recover the efficiency or the normalized SHG efficiency, therefore the variation in normalized efficiency may be caused by the formation of temperature gradient inside the crystal, arising from the self-heating. It is important to note that the normalized conversion efficiency, representing the property of the nonlinear crystal, remains stable after the first 30 hours of operation. With the normalized conversion remaining stable, we suspect that the drop in blue-to-DUV efficiency is caused by the increase in the optical loss of the cavity. We have not been able to separate the source of the loss; if it is caused by the bulk property of BBO, surface degradation, or any of the optics of the resonator. After the drop, the output power or the efficiency usually recover only partially by translating the crystal, suggesting multiple factors are involved.

Using the DUV laser source, we demonstrate MOT of Cd atoms on the $^1S_0$ - $^1P_1$ transition at 229 nm. In order to avoid the long-term degradation of the DUV power observed in Fig. 4, we operated the laser at a reduced DUV output power of approximately 80 mW. At this output power levels, 2 W of fundamental power from a commercially available tapered amplifier (TA) suffices. The TA is seeded by an external cavity diode laser that is stabilized to the reference cavity to reduce the frequency fluctuation and drift.

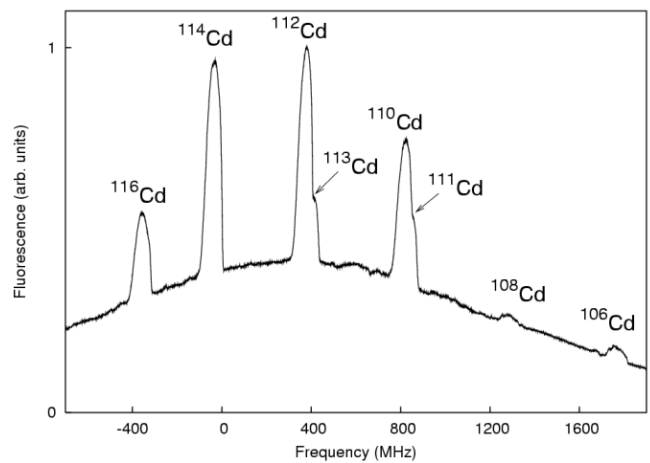

Fig. 5. Fluorescence from magneto-optically trapped Cd atoms on the $^1S_0$ - $^1P_1$ transition at 229 nm. MOT peaks are labeled with the corresponding isotopes. Broad background shows a Doppler-broadened fluorescence from the Cd vapor.

To obtain sufficient restoring force on this broad MOT transition with a natural linewidth of $\Gamma/(2\pi)$ = 90.9 MHz [8], we applied magnetic field gradient of 17 mT/cm (along the axial direction of the quadrupole magnetic field). Intensity of each MOT laser is about 0.1 $I_{sat}$, where the saturation intensity of the transition, $I_{sat}$, is 988 mW/cm$^2$. Cd atoms are supplied from a dispenser with its typical vapor pressure of 1 × 10$^{-7}$ Pa.

Figure 5 shows fluorescence from the magneto-optically trapped Cd isotopes as a function of MOT laser frequency in respect to $^{114}$Cd. The laser frequency is scanned from the lower to higher frequencies. We have trapped all stable Cd isotopes, including the Fermionic isotopes $^{113}$Cd and $^{111}$Cd, which were not observed previously [2].

In summary, we have developed a continuous-wave DUV laser source at 229 nm for the laser cooling of Cd atoms. The DUV light was obtained by the 4$^{th}$ harmonic of an OPSL as the fundamental source. A maximum output power of 0.56 W at 229 nm was observed with the blue to DUV conversion efficiency greater than 34%. The DUV source was tested for a period of 120 hours cumulative operation starting at 0.4 W, after which the power dropped to 0.2 W. We believe the main cause of this drop is the increase in optical loss in the doubling cavity. We observed indications of self-heating, in the power input-output characteristics as well as in the transmission waveform from the cavity with the PZT scanned. However, for a laser sources that will be used for much longer than 100 hours at similar output power levels, some of the parameters, including the spot size in BBO may need to be changed. Using this DUV laser source, we demonstrated MOT of all stable Cd isotopes including $^{113}$Cd and $^{111}$Cd isotopes, which are applicable to Cd-based optical lattice clocks [9].

**References**

1. S. R. J. Brueck, Proc. IEEE, **93**, 1704 (2005).
2. K. –A. Brickman, M. –S. Chang, M. Acton, A. Chew, D. Matsukevich, P. C. Haljan, V. S. Bagnato, and C. Monroe, Phys. Rev. A **76**, 043411 (2007).
3. K. Yamanaka, N. Ohmae, I. Ushijima, M. Takamoto, and H. Katori, Phys. Rev. Lett. **114**, 230801 (2015).
4. M. Kuznetsov, F. Hakimi, R. Sprague, and A. Mooradian, IEEE Photo. Tech. Lett. **9**, 1063 (1997).
5. Y. Kaneda, J. M. Yarborough, L. Li, N. Peyghambarian, L. Fan, C. Hessenius, M. Fallahi, J. Hader, J. V. Moloney, Y. Honda, M. Nishioka, Y. Shimizu, K. Miyazono, H. Shimatani, M. Yoshimura, Y. Mori, Y. Kitaoka, and T. Sasaki, Opt. Lett. **33**, 1705 (2008).
6. H. Wada, M. Oka, K. Tatsuki, M. Saito, and S. Kubota, Jpn. J. of Appl. Phys. **43,** L393-L395 (2004).
7. K. Kato, N. Umemura, and T. Mikami, Proc. SPIE **7582,** 75821L (2010).
8. H. L. Xu, A. Persson, S. Svanberg, K. Blagoev, G. Malcheva, V. Pentchev, E. Biémont, J. Campos, M. Ortiz, and R. Mayo, Phys. Rev. A **70**, 042508 (2004).
9. H. Katori, Nat. Photonics **5**, 203 (2011).

**Reference (with titles)**

1. S. R. J. Brueck, "Optical and Interferometric Lithography – Nanotechnology Enablers", Proc. IEEE, **93**, 1704 (2005).
2. K. –A. Brickman, M. –S. Chang, M. Acton, A. Chew, D. Matsukevich, P. C. Haljan, V. S. Bagnato, and C. Monroe, "Magneto-optical trapping of cadmium", Phys. Rev. A **76**, 043411 (2007).
3. K. Yamanaka, N. Ohmae, I. Ushijima, M. Takamoto, and H. Katori, "Frequency Ratio of $^{119}$Hg and $^{87}$Sr Optical Lattice Clocks beyond the SI Limit", Phys. Rev. Lett. **114**, 230801 (2015).
4. M. Kuznetsov, F. Hakimi, R. Sprague, and A. Mooradian, "High-power (>0.5-W CW) diode-pumped vertical-external-cavity surface-emitting semiconductor lasers with circular $TEM_{00}$ beams", IEEE Photo. Tech. Lett. **9**, 1063 (1997).
5. Y. Kaneda, J. M. Yarborough, L. Li, N. Peyghambarian, L. Fan, C. Hessenius, M. Fallahi, J. Hader, J. V. Moloney, Y. Honda, M. Nishioka, Y. Shimizu, K. Miyazono, H. Shimatani, M. Yoshimura, Y. Mori, Y. Kitaoka, and T. Sasaki, "Continuous-wave all-solid-state 244 nm deep-ultraviolet laser source by fourth-harmonic generation of an optically pumped semiconductor laser using $CsLiB_6O_{10}$ in an external resonator", Opt. Lett. **33**, 1705 (2008).
6. H. Wada, M. Oka, K. Tatsuki, M. Saito, and S. Kubota, "Measurement and analysis of cavity loss of a 266 nm continuous-wave solid-state laser," Jpn. J. of Appl. Phys. **43,** L393-L395 (2004).
7. K. Kato, N. Umemura, and T. Mikami, "Sellmeier and thermo-optic dispersion formulas for β-$BaB_2O_4$ (revisited)," Proc. SPIE **7582,** 75821L (2010).
8. H. L. Xu, A. Persson, S. Svanberg, K. Blagoev, G. Malcheva, V. Pentchev, E. Biémont, J. Campos, M. Ortiz, and R. Mayo, "Radiative lifetime and transition probabilities in Cd I and Cd II", Phys. Rev. A **70**, 042508 (2004).
9. H. Katori, "Optical lattice clocks and quantum metrology", Nat. Photonics **5**, 203 (2011).